\documentclass[%
twocolumn,
nofootinbib,
 amsmath,amssymb,
 aps,
 prl,
]{revtex4-2}

\usepackage{graphicx}
\usepackage[normalem]{ulem}
\usepackage{dcolumn}   
\usepackage{bm}        

\usepackage[english]{babel}

\usepackage[dvipsnames]{xcolor}
\usepackage{hyperref}
\hypersetup{
    colorlinks = true,
    linkcolor  = Blue,
    urlcolor   = Blue,
    citecolor  = Blue
}

\usepackage[labelfont = bf]{caption}
\usepackage{subcaption}

\usepackage{ulem}
\usepackage{booktabs}  

\newcommand{\aap}{Astronomy and Astrophysics}
\newcommand{\apjl}{The Astrophysical Journal Letters}

\newcommand{\mnras}{Monthly Notices of the Royal Astronomical Society}

\newcommand{\araa}{Annual Review of Astronomy and Astrophysics}

\newcommand{\ssr}{Space Science Reviews}

\newcommand{\orcid}[1]{\href{https://orcid.org/#1}{\includegraphics[height=\fontcharht\font`\B]{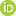}}}

\begin{document}

\title{The merger of spinning, accreting supermassive black hole binaries}


\author{Lorenzo Ennoggi\:\orcid{0000-0002-2771-5765}}
\affiliation{Center for Computational Relativity and Gravitation, Rochester Institute of Technology \& School of Physics and Astronomy, Rochester, New York 14623, USA }

\author{Manuela Campanelli\:\orcid{0000-0002-8659-6591}}
\affiliation{Center for Computational Relativity and Gravitation, Rochester Institute of Technology \& School of Physics and Astronomy, Rochester, New York 14623, USA }

\author{Julian Krolik\:\orcid{0000-0002-2995-7717}}
\affiliation{Physics and Astronomy Department, Johns Hopkins University, Baltimore, Maryland 21218, USA}

\author{Scott C. Noble\:\orcid{0000-0003-3547-8306}}
\affiliation{Gravitational Astrophysics Lab, NASA Goddard Space Flight Center, Greenbelt, Maryland 20771, USA}

\author{Yosef Zlochower\:\orcid{0000-0002-7541-6612}}
\affiliation{Center for Computational Relativity and Gravitation \& School of Mathematics and Statistics, Rochester Institute of Technology, Rochester, New York 14623, USA}

\author{Maria Chiara de Simone\:\orcid{0009-0008-8088-1392}}
\affiliation{Center for Computational Relativity and Gravitation, Rochester Institute of Technology \& School of Physics and Astronomy, Rochester, New York 14623, USA}






\date{\today}

\begin{abstract}
Because they are likely to accrete substantial amounts of interstellar gas, merging supermassive binary black holes are expected to be strong multimessenger sources, radiating gravitational waves, photons from thermal gas, and photons from relativistic electrons energized by relativistic jets. Here we report on a numerical simulation that covers the late inspiral, merger, and initial postmerger phase of such a system where both black holes have the same mass and spin, and both spin axes are parallel to the orbital angular momentum. The simulation incorporates both 3D general relativistic magnetohydrodynamics and numerical relativity. The thermal photon power during the late inspiral, merger, and immediate postmerger phases is drawn from strong shocks rather than dissipation of turbulence inside a smoothly structured accretion disk as typically found around accreting single black holes.
We find that the thermal photon and jet Poynting flux outputs are closely related in time, and we posit a mechanism that enforces this relation.
The power radiated in both photons and jets diminishes gradually as merger is approached, but jumps sharply at merger to a noisy plateau. Such a distinct lightcurve should aid efforts to identify supermassive black hole mergers, with or without accompanying gravitational wave detections.
\end{abstract}


\maketitle

\textit{Introduction---}A central goal in multimessenger astronomy is to connect gravitational wave (GW) signals from supermassive binary black hole (SMBBH) systems with their electromagnetic (EM) counterparts~\cite{Begelman1980, Colpi2014}. Photon observations are critical for identifying and characterizing these systems all the way from inspiral to merger and its aftermath~\cite{Haiman2009, Bogdanovic2022}. For example, such signals may enable the detection of SMBBHs before the GWs can be detected, alert us to recent mergers, and allow us to track their orbital evolution~\cite{Tanaka2013, Kelley2018}. The unique structure of the accreting gas---often featuring a circumbinary disk (CBD) with a central gap and individual ``minidisks''---is expected to generate distinct spectral features and periodic light curves that distinguish these systems from single accreting supermassive black holes, but there is considerable controversy over just what these features may be (e.g.,~\cite{Bogdanovic2022} for a review).

Some have argued that photon emission should drop sharply as the binary shrinks and not revive until well after merger~\cite{Pringle1991, Liu2003, MilosavljevicPhinney2005, Farris2015, Krauth2023, Krauth2025} while others predict that it might drop by a factor ${\sim O\left(1\right)}$ and then revive quickly after merger~\cite{MacFadyenMilosavljevic2008, Shi2012, Noble2012, Farris2014, Gold2014a, Gold2014b, ShiKrolik2015, Avara2024, Ennoggi2025}. Many have suggested ways the luminosity and jet Poynting flux should generically be modulated periodically~\cite{MacFadyenMilosavljevic2008, Shi2012, Noble2012, Sesana2012, Dorazio2013, Roedig2014, Kelly2017, Tang2018, Paschalidis2021, Cattorini2021, Noble2021, Bright2023, Fedrigo2023}, while some have argued that modulations should be weak or indistinct throughout the inspiral~\cite{Noble2012} or as merger is approached~\cite{Gold2014a, Bright2023}. Analytic estimates~\cite{Tanaka2013, Roedig2014} and smoothed particle hydrodynamics (SPH) simulations~\cite{Sesana2012, Farris2015} predict a significant ``notch" in the thermal spectrum well before merger, while others have argued this feature does not appear at all~\cite{Farris2015} and GRMHD simulations of late inspiral indicate the notch may disappear near merger~\cite{dAscoli2018, Gutierrez2022, Gutierrez2024, Ennoggi2025}. To date, none of these signals has been observed.

\smallskip
Using our recently developed computational tools for simultaneous solution of the magnetohydrodynamics (MHD) and Einstein field equations~\cite{Ennoggi2025}, we have found significant new results about dynamics shortly before, during, and after the merger of a SMBBH with spinning black holes, and these results have strong implications for photon emission and jet launching. Here we focus on the case in which both spin axes are parallel to the angular momentum of the binary. This simplest spin configuration is favored by evolutionary models~\cite{Bogdanovic2007, MillerKrolik2013}.

Prior to merger, both black holes can support jets. Remarkably, we find a strong linear correlation between the Poynting luminosity of the jets and the photon luminosity from the gas. Moreover, during the first portion of the inspiral we simulate, when the SMBBH separation is ${\sim\!15\text{--}20\,M}$ (the gravitational radius ${r_g\equiv GM/c^2 = M}$ and the gravitational timescale ${t_g\equiv r_g/c = M}$ for ${G=c=1}$, so both length and time are in units of $M$), both luminosities are more tightly correlated with the gas mass in which the black holes are immersed than with the mass accretion rate onto the black holes. Over the course of the inspiral, both luminosities decrease gradually by factors of several, but neither disappears; at the moment of merger, both jump back to roughly their original level. Such distinct lightcurves might be used to identify SMBBH mergers even without direct detection of GWs.

\bigskip
\textit{Calculational details---} Using the same codes and numerical techniques described in~\cite{Ennoggi2025}, we simulated numerically the evolution of an equal-mass accreting SMBBH whose black holes spin. The initial state of the gas is the same relaxed CBD state used in~\cite{Ennoggi2025} (at time ${\sim\!99200\,M}$), and, following the same ``hand-off'' steps detailed in that work, we created the initial data for a simulation that combines full numerical relativity with general relativistic MHD using the \textsc{IllinoisGRMHD} code~\cite{Etienne2015, Werneck2023}. However, unlike the previous work in which the black holes had no spin, at the time of hand-off we set up a two-puncture spacetime~\cite{Campanelli2006, Baker2006} where both black holes have dimensionless spin parameter 0.8 and both spin axes are parallel to the angular momentum of the binary. Spins have a negligible effect on the relaxed disk state~\cite{LopezArmengol2021}, so we can safely use the nonspinning relaxed CBD for a spinning binary. Because in the moving puncture gauge the horizon's radius decreases with increasing spin, we added an additional refinement level to the computational grid with radius ${0.5\,M}$ and resolution ${M/128}$ around each black hole and enlarged the radius of the next coarser level from ${0.5\,M}$ to ${1.5\,M}$. These changes ensure that the horizons lie well within the finest refinement level and are covered by ${\sim\!80}$ (${\sim\!64}$) points across their largest (smallest) dimension.

\begin{figure*}
    \centering
    \includegraphics[width = \textwidth]{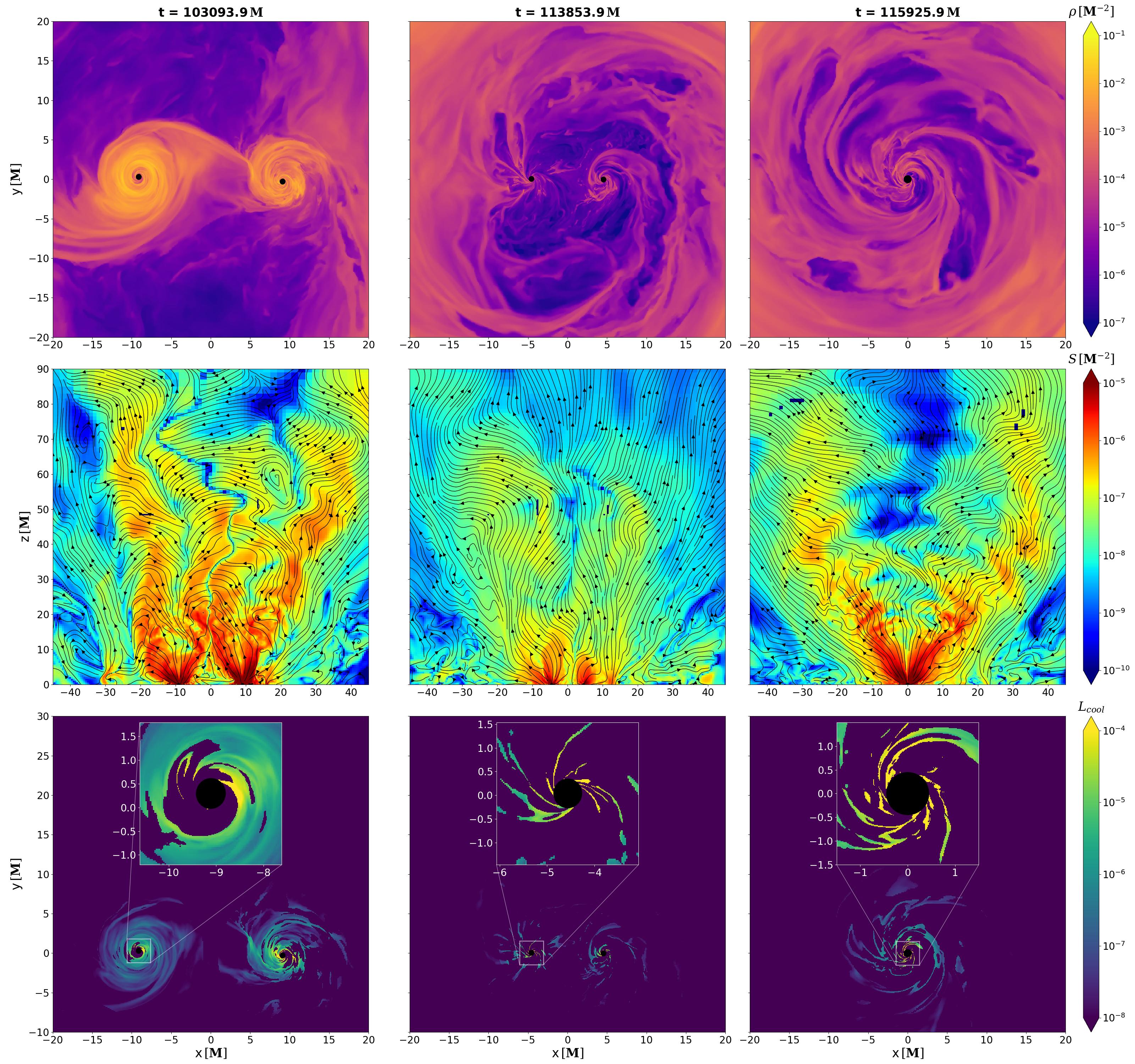}
    \captionsetup{justification = raggedright} 
    \caption{Mass density on equatorial slices (top panels), magnitude of the Poynting vector and magnetic field-lines on vertical slices (middle panels), and photon emissivity on equatorial slices (bottom panels).  Three times are shown for all quantities: early inspiral (left panels),shortly before merger (middle panels), and shortly after merger (right panels).  At both pre-merger times, the black holes lie close to the {$x$-axis.}}
    \label{fig: snapshots}
\end{figure*}

\bigskip
\textit{Principal results---}As the inspiral begins, minidisks quickly form around each black hole (top-left panel of Figure~\ref{fig: snapshots}). As in~\cite{Ennoggi2025}, soon thereafter each minidisk begins to alternate between a ``disk-like'' state, with maximum mass and minimum accretion rate onto its black hole, and a ``stream-like'' state, with opposite properties~\cite{Avara2024}. However, the initial time-averaged minidisks' masses are ${4\text{--}5\times}$ larger than in the nonspinning configuration of~\cite{Ennoggi2025}, mostly due to the smaller radius of the innermost stable circular orbit (ISCO) of a spinning black hole. The minidisks also retain their mass for longer: the mass of one takes a time ${\sim\!4300\,M}$ to fall to a value matching the initial minidisk mass in the nonspinning run, and the other does not fall to that level for another ${\sim\!3000\,M}$. This extended lifetime of the minidisks is a consequence of the ``hangup'' effect~\cite{Campanelli2006HangUp}, in which repulsive spin-orbit coupling both lengthens the inspiral and holds the constituent gas at larger radii before the merger~\cite{LopezArmengol2021}.

While they last, the minidisks exchange mass through ``sloshing''~\cite{Bowen2017}. At around ${t\simeq 112500\,M}$, the minidisks dissolve. Thereafter, rather than being distributed through the minidisks, gas heating is highly concentrated in narrow fronts (see Figure~\ref{fig: snapshots}), a strong signal that the heating in this stage is dominated by shocks.

Accretion carries magnetic flux to the horizons, enabling each black hole to drive a relativistic and Poynting-flux dominated jet along its spin axis. The center-left panel of Figure~\ref{fig: snapshots} shows that the maximum magnitude of the Poynting vector in the bulk of the jets is ${\sim\!10^{-4}M^{-2}\text{--}10^{-3}M^{-2}}$, comparable to what was found in~\cite{Combi2022} for a spin 0.6 binary. As already observed in, e.g.,~\cite{Ressler2024}, in the early inspiral the jets are separated by a layer of low-magnetization plasma out to a distance ${80\text{--}100\,M}$ from the orbital plane, where the two jets begin merging into a single structure. Because the black hole spins are parallel and both holes accrete magnetic field with the same mean polarity, near the orbital axis the azimuthal magnetic fields from the two black holes are oppositely directed and magnetic reconnection can occur.

\begin{figure*}
    \centering
    \includegraphics[width = \textwidth]{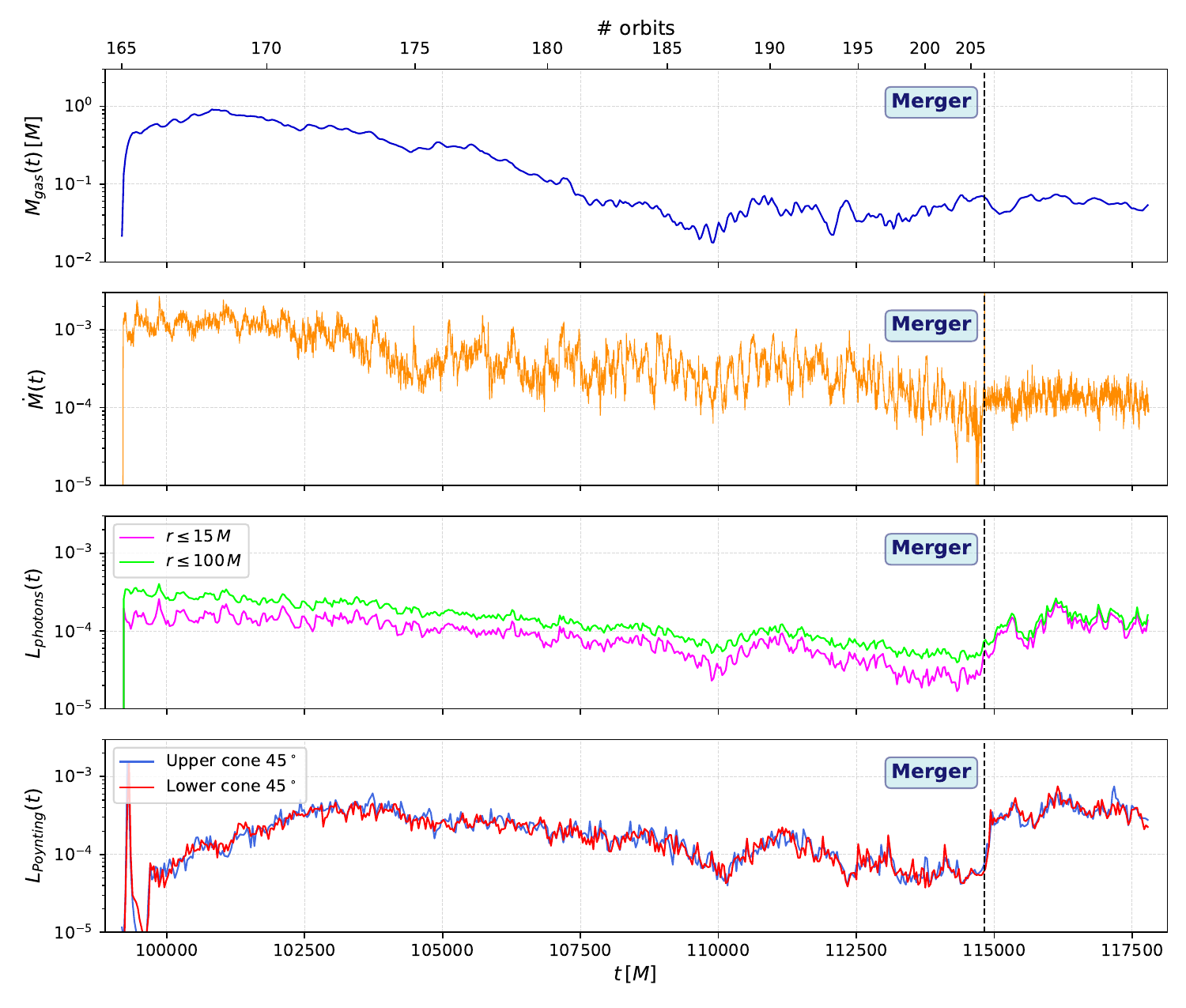}
    \captionsetup{justification = raggedright} 
    \caption{Mass within ${r\leq 15\,M}$ (top panel), total accretion rate onto the binary (second panel), photon luminosity within ${r\leq 15\,M}$ and ${r\leq 100\,M}$ (third panel), and Poynting flux in the jet regions at ${r=100\,M}$ (bottom panel).}
    \label{fig: plots}
\end{figure*}

\smallskip
The volume-integrated cooling rate is the bolometric photon luminosity. Because the minidisks are more massive than those in~\cite{Ennoggi2025}, the luminosity from the binary region ${r\leq 15\,M}$ (Figure~\ref{fig: plots}) is initially ${\sim\!50\%}$ higher than its equivalent in~\cite{Ennoggi2025}. On the other hand, and unsurprisingly, the luminosity from the rest of the cavity and the bulk of the CBD (i.e., the spherical shell ${15\,M\leq r\leq 100\,M}$) is comparable to that of~\cite{Ennoggi2025}. Taken all together, the radiation from the inner cavity region (${r\leq 15\,M}$) during the inspiral is generally ${\gtrsim 0.6}$ of the total; after merger, it is essentially all of the total luminosity.

As Figure~\ref{fig: plots} shows, both the photon luminosity and the Poynting flux have overall downward trends during the inspiral, although there are also temporary recoveries lasting as long as ${\sim\!2000\,M}$. Qualitatively, the photon lightcurve in the spinning case resembles the lightcurve in the nonspinning case. However, there is a significant quantitative contrast: the degree of decline in the spinning case for the luminosity of the innermost region (${r\leq 15\,M}$) is considerably smaller than in the nonspinning case---a factor ${\sim\!4}$ as opposed to a factor ${\sim\!8}$.

\begin{figure}
    \centering
    \includegraphics[width = \linewidth]{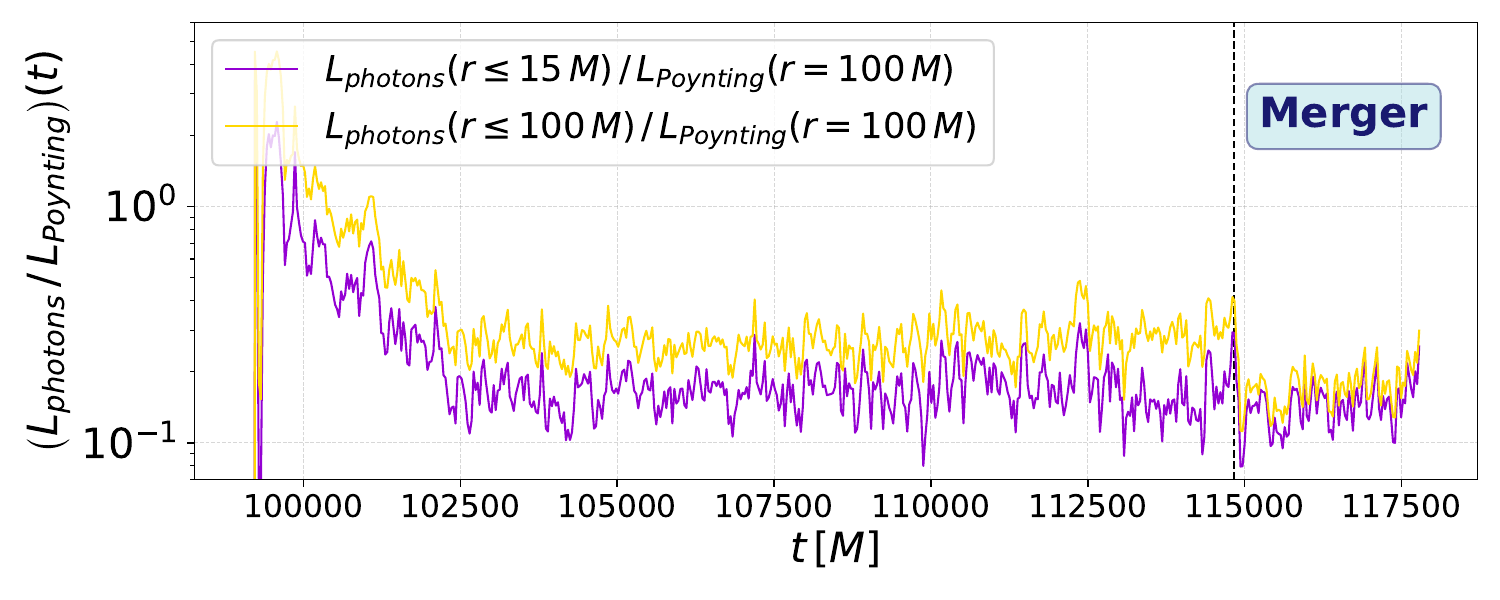}
    \captionsetup{justification = raggedright} 
    \caption{Ratio of ${L_\text{photons}}$ from ${r\leq 15\,M}$ (purple) or ${r\leq 100\,M}$ (yellow) to ${L_\text{Poynting}\left(r = 100\,M\right)}$.
    The averages for ${102500\,M\lesssim t\lesssim 110000\,M}$ are $0.164$ (photons from ${r\leq 15\,M}$) and $0.262$ (${r\leq 100\,M}$).}
    \label{fig: L_photons over L_Poynting}
\end{figure}

The most striking---and surprising---feature of the spinning merger's power output is that the time dependence of Poynting flux in the jet very strongly resembles the photon lightcurve. To demonstrate just how closely they follow each other, in Figure~\ref{fig: L_photons over L_Poynting} we show the ratio of the two lightcurves. From ${t\simeq 102500\,M}$ until the merger at ${t\simeq 115000\,M}$, this ratio shows almost no variation on timescales longer than a few hundred $M$: ${L_\text{photons}\left(r\leq 100\,M\right)/L_\text{Poynting}\approx 0.26\pm 0.04}$\,. Nothing of the sort is seen in the nonspinning case because spinless black holes are almost wholly incapable of supporting Poynting flux.

In conventional accretion flows, a similar relationship between ${L_\text{photons}}$ and ${L_\text{Poynting}}$ would be expected because the radiative efficiency (the ratio of power output to accretion rate) for both photon radiation~\cite{NovikovThorne1973, ShakuraSunyaev1973, NobleKrolikHawley2009} and Poynting power in a jet~\cite{HawleyKrolik2006, DavisTchekhovskoy2020} is constant when averaged over short-timescale fluctuations. However, in this context the relationships are quite different, and change with time (see details in the Supplemental Material~\cite{SupplementalMaterial}). Three distinct periods can be discerned: the early inspiral (${103000\,M\lesssim t\lesssim 108000\,M}$), the late inspiral (${108000\,M\lesssim t\leq t_\text{merger}}$), and the postmerger phase (${t > t_\text{merger}}$)\,. The photon power is moderately well correlated with the accretion rate throughout the inspiral (Pearson ${r\simeq 0.57}$ in the early inspiral, ${r\simeq 0.76}$ in the late inspiral), but, as illustrated in Figure~\ref{fig: plots}, during the early inspiral is even better correlated with the mass in the inner cavity, ${M_\text{gas}\left(r\leq 15\,M\right)}$: ${r\simeq 0.83}$\,. Moreover, the best-fit scaling with ${M_\text{gas}}$ is ${L_\text{photons}\propto M_\text{gas}^{0.32}\left(r\leq 15\,M\right)}$. After the merger, ${L_\text{photons}}$ has essentially no correlation with either ${\dot{M}}$ or ${M_\text{gas}\left(r\leq 15\,M\right)}$. The relations between ${L_\text{Poynting}}$ and accretion rate and inner gas mass are similar, but with somewhat weaker correlations. This behavior immediately signals that the radiation mechanisms for both photons and Poynting flux are qualitatively different from those of ordinary black hole accretion systems.

The tight link between ${L_\text{photons}}$ and ${L_\text{Poynting}}$ can be understood if the gas is in thermal balance with a cooling time ${t_\text{cool}}$ independent of the heating rate, and the magnetic flux on the black holes is confined by the thermal pressure of surrounding gas rather than the ram pressure of accreting mass. The surrounding gas should be in thermal balance
because, independent of temperature, the cooling time is comparable to the orbital dynamical time, and the inflow time (${M_\text{gas}/\dot{M}}$) is generally ${\gtrsim 10\times}$ longer. The Poynting luminosity ${L_\text{Poynting}\!\sim\!B^2 M^2 f\!\left(a/M\right)}$\,, where $B$ is the poloidal magnetic field on the horizon, the surface area of the black hole is ${\sim\!M^2}$, and ${f\!\left(a/M\right)}$ describes how the Poynting luminosity depends on the spin parameter ${a/M}$~\cite{McKinneyGammie2004}.
Thermal balance in the gas implies that the gas pressure ${P\!\sim\!L_\text{photons}t_\text{cool}/r^3}$ because the cooling rate is exactly the photon luminosity; if the magnetic pressure $B^2$ is matched by $P$,
it immediately follows that ${L_\text{Poynting}\propto L_\text{photons}}$\,.

\smallskip
There are plausible physical circumstances in which ${t_\text{cool}}$ is independent of ${L_\text{photons}}$\,, such as a roughly round density distribution whose opacity (e.g., electron scattering) is independent of temperature. However, even for more general circumstances, magnetic confinement by thermal pressure leads to a tight relation between ${L_\text{photons}}$ and ${L_\text{Poynting}}$\,; all that changes is the logarithmic slope of that relation.

That the two luminosities are ${\propto\!M_\text{gas}^{1/2}\left(r\leq 15\,M\right)}$ requires that the heating rate per unit mass be ${\propto\!M_\text{gas}^{-1/2}\left(r\leq 15\,M\right)}$. This might be the case in the early inspiral if shocks between gas streams in the minidisks are restrained when the minidisks' masses are relatively large.

\smallskip
As in the nonspinning case~\cite{Ennoggi2025}, both ${M_\text{gas}\left(r\leq 15\,M\right)}$ and ${\dot{M}}$ are essentially unchanged as a result of the merger, even as the luminosity of the inner region increases sharply---in both the spinning and nonspinning cases by a factor ${\sim 5}$.
For several thousand $M$ after merger, both ${L_\text{photons}}$ and ${L_\text{Poynting}}$ vary widely with little correlation to either ${\dot{M}}$ or ${M_\text{gas}\left(r\leq 15\,M\right)}$, but without any overall trend. Moreover, the radiative efficiency ${L_\text{photons}/\dot{M}\!\sim\!1\text{--}2}$,
an order of magnitude greater than the level usually envisaged in relativistic accretion. Taken together, these facts suggest that the heating processes during this period are relativistic shocks (see Figure~\ref{fig: snapshots}) passing through a highly irregular density distribution.

On the other hand, the luminosity from the rest of the cavity and the bulk of the CBD is largely unaffected by the merger. For this reason, the total luminosity from within ${100\,M}$ of the center of mass increases by a factor ${\sim\!2}$ in the present case even while the central portion increases by a factor ${\sim\!5}$.

\bigskip
\textit{Conclusions}---We have shown that, contrary to work based on Newtonian 2D hydrodynamics~\cite{Krauth2023} that predicts a long period of very low luminosity starting immediately prior to a SMBBH merger and extending long after it, significant photon luminosity and jet power persist throughout such an event, and both sorts of luminosities rise to their previous levels immediately upon the merger taking place. Although minidisks disappear during the late inspiral, removing the possibility of heat production through dissipation of MHD turbulence, shocks between disorganized gas streams replace this mechanism, especially in the postmerger regime.

Particularly if the Poynting luminosity lightcurve is mirrored in the lightcurve of photons radiated by electrons accelerated by the jet, these findings about lightcurves during the late inspiral and immediate postmerger phase of a SMBBH merger could have important observational implications. A gradual decrease in luminosity radiated by the accreting gas along with a parallel decrease in the synchrotron or inverse Compton emission associated with the jets, followed by a quick uptick in both would be an extremely distinct signature of a SMBBH merger. Moreover, the timescales for these variations, ${\sim\!10^4 M\simeq 2\,M_8}$~months for the decline and ${\sim\!100\,M\simeq 0.5\,M_8}$~days for the sharp rise, are well within humanly accessible timescales for black hole masses ${\sim\!10^7\text{--}10^9 M_\odot}$\,. In addition, as already noted in~\cite{Ennoggi2025}, because the location from which much of the light is emitted changes sharply at merger, an equally sharp change in the light's spectral properties should be expected.

Identification of such a signal could be useful for a number of purposes. It would provide the first clear indication that mergers of supermassive black holes actually occur, vindicating the expectation raised by many studies of the cosmological evolution of galaxies that these events happen. If more than a handful were identified within the next decade, it would provide explicit guidance about the population of mergers potentially detectable by spaceborne GW observatories like LISA. Whenever such an observatory does begin operations, searching for these lightcurves would be an immensely useful tool for the localization of the GW events detected by that observatory. Lastly, the close coupling of photon and Poynting luminosity is a demonstration of the fluid-magnetic coupling responsible for confining magnetic flux on the event horizons of black holes.

\bigskip
\textit{Acknowledgments---}This work was supported by NASA Theory and Computational Astrophysics Network grant 80NSSC24K0100. M.C. and Y.Z. acknowledge support from NSF awards AST-2009330, AST-1516150, PHY-2110338, and PHY-1707946. J.K. was partially supported by NSF grants AST-2009260 and PHY-2110339. We are also grateful to our collaborators Jay V. Kalinani, Michael Chabanov, Carlos O. Lousto, Vassilios Mewes, Liwei Ji, Luciano Combi and Jeremy Schnittman for valuable discussions and comments.

Computational resources were provided by TACC's Frontera supercomputer allocations PHY-20010 and AST-20021. Additional resources were provided by RIT's BlueSky, Green Prairies, and Lagoon clusters, acquired with NSF grants PHY-2018420, PHY-0722703, PHY-1229173 and PHY-1726215.

\bibliographystyle{apsrev4-2}
\bibliography{PRLBibliography}  

\end{document}